\documentclass{article}
\usepackage{spconf,amsmath,graphicx}
\usepackage{booktabs}
\usepackage{subcaption}

\DeclareMathOperator*{\argmax}{arg\,max}

\title{Controlling Whisper: Universal Acoustic Adversarial Attacks to Control Multi-task Automatic Speech Recognition Models}
\twoauthors
 {Vyas Raina}
	{University of Cambridge\\
	vr313@cam.ac.uk}
 {Mark Gales}
	{University of Cambridge\\
	mjfg100@cam.ac.uk}

\begin{document}
%
\maketitle
\begin{abstract}
Speech enabled foundation models, either in the form of flexible speech recognition based systems or audio-prompted large language models (LLMs), are becoming increasingly popular. One of the interesting aspects of these models is their ability to perform tasks other than automatic speech recognition (ASR) using an appropriate prompt. For example, the OpenAI Whisper model can perform both speech transcription and speech translation. With the development of audio-prompted LLMs there is the potential for even greater control options. In this work we demonstrate that with this greater flexibility the systems can be susceptible to model-control adversarial attacks. Without any access to the model prompt it is possible to modify the behaviour of the system by appropriately changing the audio input. To illustrate this risk, we demonstrate that it is possible to prepend a short universal adversarial acoustic segment to any input speech signal to override the prompt setting of an ASR foundation model. Specifically, we successfully use a universal adversarial acoustic segment to control Whisper to always perform speech translation, despite being set to perform speech transcription. Overall, this work demonstrates a new form of adversarial attack on multi-tasking speech enabled foundation models that needs to be considered prior to the deployment of this form of model.
\end{abstract}
\begin{keywords}
ASR, Adversarial Attacks, Control
\end{keywords}

\section{Introduction}

\begin{figure}[htb!]
    \centering
    \includegraphics[width=\linewidth]{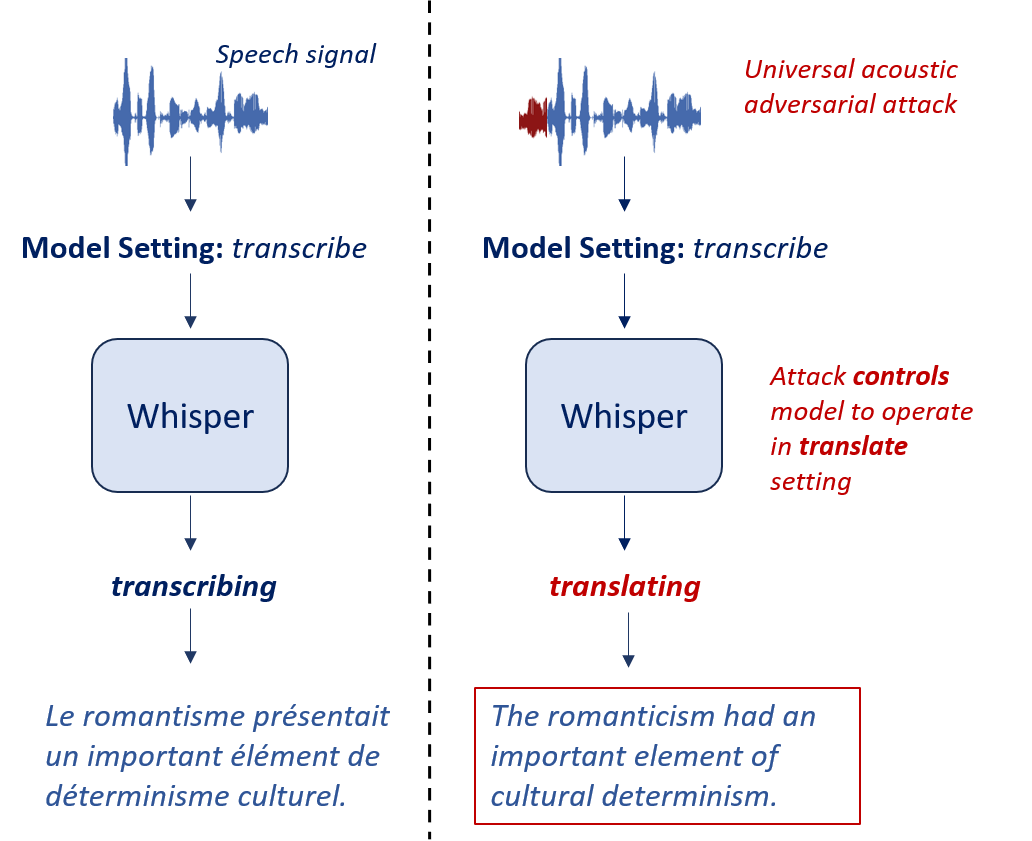}
    \caption{A short universal acoustic adversarial segment can be prepended to any input speech signal to \textit{control} the behavior of a multi-task Automatic Speech Recognition (ASR) model. For example, Whisper's \textit{transcribe} setting can be overridden such that it operates in its \textit{translate} setting.}
    \label{fig:main}
\end{figure}
 In the form of flexible automatic speech recognition (ASR) systems~\cite{radford2022robust} or audio-prompted Large Language Models (LLMs)~\cite{tang2024salmonn}, speech enabled foundation models are increasingly able to perform other speech processing tasks beyond just speech transcription - we refer to such systems as multi-task ASR models. These models are trained to execute a diverse set of speech processing tasks within a single framework, enabling their deployment across a wide range of applications. A notable example of such a model is OpenAI's Whisper~\cite{radford2022robust}, which employs an encoder-decoder Transformer architecture to perform both speech transcription and speech translation. The task to be performed by Whisper is determined by a `task tag' included in the textual prompt to the decoder. It is anticipated that future advancements will lead to the development of increasingly flexible speech-enabled foundation models capable of performing a broader number of speech processing tasks withing a single framework.

However, in this work we are the first to demonstrate that the capability of ASR models to perform multiple tasks introduces a new class of vulnerabilities: model-control adversarial attacks. These attacks aim to manipulate a model's behavior such that, despite being configured for a specific task in deployment, the adversary can override the task setting and induce the model to perform an alternative target task (the target task is required to be within the set of tasks the model has been trained to be able to do). Historically, adversarial attacks on ASR models have primarily focused on disrupting the model's performance or causing it to transcribe a predetermined phrase~\cite{DBLP:journals/corr/abs-1711-03280, cisse2017houdini}. In contrast, our proposed model-control adversarial attacks have a distinctly different objective where they instead seek to override the operational setting of a multi-task ASR model, forcing it to function in an unintended mode.

To highlight the susceptibility of multi-task ASR models to such control-attacks, this work proposes a practical method to subvert the `transcription' setting of the Whisper model, and instead encouraging Whisper to execute speech `translation'. Given that an adversary typically does not have access to the textual decoder prompt, we adopt a threat model where the attack is restricted to the acoustic space. Our findings reveal that a short ($\sim$5 seconds) universal adversarial acoustic segment can be prepended to any speech signal and override Whisper's `transcribe' setting and execute the `translate' function instead (depicted in Figure \ref{fig:main}). Experimental evaluations across four languages indicate that this universal acoustic attack segment can consistently manipulate Whisper's behavior for nearly all test samples. Further examination of more constrained attack scenarios reveals an interesting bi-modal pattern: the attacks are either highly successful in overriding the command or entirely ineffective for specific samples.

Overall, this work illustrates the vulnerability of flexible multi-task speech foundation systems to a novel form of adversarial attack. Model-control attacks exploit the model's capability to perform various tasks, thereby enabling the adversary to override the intended operational setting. As ASR models become increasingly flexible and capable of handling more tasks, it is crucial to consider the potential risks posed by model-control adversarial attacks when deploying these models in real-world applications.

\section{Related Work}

\subsection{Multi-task Speech Foundation Models}

Speech enabled foundation models are trained on a large quantity of data to process input speech signals and perform various different speech processing tasks. The development of these models have taken two main forms. First, several approaches have attempted to extend powerful textual decoder-only Large Language Models (LLMs) to support direct speech inputs with a trained connection module to provide the embedded audio as a soft prompt~\cite{tang2024salmonn, huang2024dynamicsuperb, chen2023xllm, wu2023decoderonly, yu2023connecting, fathullah2023prompting}. Alternatively, generalist speech enabled foundation models have emerged in the form of flexible Automatic Speech Recognition (ASR) models~\cite{radford2022robust} using an encoder-decoder architecture~\cite{NIPS2017-3f5ee243}. ASR models are traditionally designed to perform a single task: transcribe the input speech signal. However, recent powerful ASR models have been augmented with further speech processing abilities. Some of the most powerful flexible ASR models include OpenAI's Whisper model~\cite{radford2022robust} and NVIDIA's Canary model~\cite{Rastorgueva-Koluguri-2024}, where these models can perform both speech transcription and speech translation tasks. These models adopt an encoder-decoder architecture, and a textual task tag is input to the decoder to indicate the speech task to be performed (transcribe or translate). Given its popularity, this work uses Whisper to illustrate the risk of control-attacks on such multi-task ASR models.

\subsection{Audio Adversarial Attacks}

Traditional audio adversarial attacks on ASR models have one of two objectives: corrupt the output transcription or deceive the ASR model into transcribing a desired target phrase. Early research on audio attacks explored gradient-based methods to perturb input audio for end-to-end ASR systems like WaveCNN and HMM-DNN, aiming to increase word error rates (WER) in transcriptions~\cite{DBLP:journals/corr/abs-1711-03280, cisse2017houdini}. Later studies introduced targeted attacks on ASR systems such as DeepSpeech, HMM-DNN, and LSTM-based neural networks to generate specific transcriptions~\cite{yuan2018commandersong, DBLP:journals/corr/abs-1801-01944, DBLP:journals/corr/abs-1805-11852, qin2019imperceptible}, while other research focused on making audio adversarial attacks more imperceptible~\cite{schönherr2018adversarial, Schnherr2018AdversarialAA}. Practical advancements include generating universal adversarial perturbations for various speech signals, although initial methods required synchronizing with the entire speech signal~\cite{DBLP:journals/corr/abs-1905-03828}. This was addressed by creating perturbations that do not need to be synchronized with the source speech signal~\cite{10.1145-3372297.3423348}, and extending these attacks to newer end-to-end ASR systems like LAS, CTC, and RNN-T~\cite{lu2021exploring}. Additional techniques involve transferability from substitute models~\cite{247642, 9348184, ma2021simulating}, evolutionary attacks~\cite{DBLP:journals/corr/abs-1801-00554, khare2019adversarial, taori2019targeted, du2019sirenattack, Zheng_2021}, utterance-based attacks~\cite{raina_gales_knill_2020}, and featurization attacks~\cite{197215, DBLP:journals/corr/abs-1708-09537, abdullah2019practical}. With the establishment of the Whisper model, new vulnerabilities to audio adversarial attacks have been identified, showing for example that adversarial attacks can lead to incorrect transcriptions~\cite{olivier2023kind} or even force entirely muted outputs for any speech input~\cite{raina2024muting}.

However, with the emergence of the multi-task speech enabled foundation models, in this work, we demonstrate that there is the threat of a new kind of adversarial attack: model-control attack. With the flexibility to perform multiple tasks, an adversary can override the operational setting of a deployed ASR model and force it to perform a different target task - i.e., take control of the model. With Whisper, we illustrate that despite being deployed in `transcribe' mode, a simple universal adversarial attack can always force Whisper to operate in `translate' mode.

\section{Multi-task Automatic Speech Recognition Models}
This section describes how encoder-decoder Automatic Speech Recognition (ASR) systems, such as Whisper, can perform multiple speech processing tasks. Continuous-time speech is sampled to create a sequence of samples, $\mathbf{x}=x_{1:N}$ of $N$ frames. An ASR system maps this sampled speech/audio signal, $\mathbf{x}$, to generate text, $\mathbf{y} = y_{1:M}$, traditionally for speech transcription. However, this generated text can also serve other speech processing tasks, such as speech translation. Whisper, a flexible multi-task ASR system, uses an encoder-decoder architecture with parameters $\theta$ to auto-regressively generate the output text tokens. The likelihood of an output sequence $\mathbf{y}$ is modeled as:
\begin{equation}
P(\mathbf{y} | \mathbf{x}, \mathcal{T}) = \prod_m P(y_m | y_{<m}, \mathbf{x}, \mathcal{T}; \theta),
\end{equation}
where $\mathcal{T}$ defines the speech processing task, and \newline $P(y_m | y_{<m}, \mathbf{x}, \mathcal{T}; \theta)$ is the probability predicted by the decoder for output token $y_m$, given $\mathbf{x}$ at the encoder input and $y_{<m}$ as the previously decoded tokens passed to the decoder input. To enable Whisper to perform multiple speech processing tasks, the task $\mathcal{T}$ is set via special text tokens at the decoder input. The first token input to the decoder is set as \texttt{<|startoftranscript|>}, followed by a token to indicate the source audio language, for example, \texttt{<fr>} for French. The next special token sets the task, such as \texttt{<transcribe>} or \texttt{<translate>}. With these special tokens in the decoder history, Whisper can flexibly perform either speech transcription ($\mathcal{T}=\texttt{tc}$) or speech translation ($\mathcal{T}=\texttt{tl}$) based on the specified task token.

\section{Model-control Adversarial Attacks} \label{sec:attack}
The objective of a model-control adversarial attack is to override the operational task setting of a multi-task ASR model, such that the model performs an alternate task as desired by the adversary. To illustrate this form of attack, in this section we propose a practical method to override Whisper's transcription setting and force Whisper instead to perform speech translation - this is depicted in Figure \ref{fig:main}.

\subsection{Threat Model}
When Whisper is deployed for a specific speech processing task, an adversary does not have access to the internal structure of the model and cannot simply change the special tokens input to the decoder to alter the task setting. Realistically, an adversary can only modify the source audio to achieve their goal of overriding Whisper's task setting. Thus, we assume the adversary can only make changes in the acoustic space. Moreover, our threat model requires the adversarial attack to be \textit{easy} to apply, as complex manipulations of the audio are impractical for live speech processing. To address this, we use a \textit{prepend} adversarial attack form, where the attack requires only a short acoustic adversarial segment to be prepended to the input speech. Learning a different adversarial manipulation for each new speech signal is impractical, especially for live streamed audio. Therefore, our threat model aims to learn a universal audio manipulation—a single short acoustic adversarial segment that can be prepended to any speech signal to achieve the desired model control. While we allow whitebox access (gradient access to the model) for learning the universal attack segment, it must be universally applicable to unseen speech signals.

\subsection{Attack Method} \label{sec:att-method}
To perturb a speech signal $\mathbf x = x_{1:N}$, we prepend a short adversarial audio segment of $T$ frames, $\tilde{\mathbf x} = \tilde{x}_{1:T}$. The perturbed speech signal becomes $\tilde{\mathbf{x}} \oplus \mathbf{x}$, where $\oplus$ represents concatenation in the raw audio space. To override Whisper's `transcribe' setting with `translate,' we need to find the optimal adversarial audio segment, $\hat{\tilde{\mathbf{x}}}$, that maximizes the likelihood of generating the translated sequence, despite Whisper running in transcribe mode:
\begin{align} \label{eqn:single}
    \hat{\tilde{\mathbf x}} &= \argmax_{\tilde{\mathbf x}} P(\mathbf{y}^{(\texttt{tl})} | \tilde{\mathbf x}\oplus\mathbf x, \mathcal{T}=\texttt{tc}),\\
    \mathbf{y}^{(\texttt{tl})} &= \argmax_\mathbf{y} P(\mathbf{y} | \mathbf x, \mathcal{T}=\texttt{tl}).\nonumber
\end{align}
To learn a \textit{universal} acoustic attack, we adapt Equation \ref{eqn:single} to maximize the likelihood of generating the translated sequence over a training set of samples, $\{\mathbf x_j\}_{j=1}^J$,
\begin{equation} \label{eqn:universal}
    \hat{\tilde{\mathbf x}} = \argmax_{\tilde{\mathbf x}} \prod_j P(\mathbf{y_j}^{(\texttt{tl})} | \tilde{\mathbf x}\oplus\mathbf x_j, \mathcal{T}=\texttt{tc}).
\end{equation}
To maximize the likelihood of Equation \ref{eqn:universal}, standard gradient descent based approaches can be used to update $\tilde{\mathbf x}$. It is important for the adversarial audio segment generated to be somewhat imperceptible such that it is not flagged as suspicious when prepended to natural speech signals. We achieve this imperceptibility in two ways: by ensuring the adversarial audio segment is short in duration and by limiting its `power' or amplitude relative to natural speech. To limit the power, we introduce a constraint in the optimization objective of Equation \ref{eqn:universal} that limits the amplitude of the adversarial audio,
\begin{equation} \label{eqn:const}
    ||\hat{\tilde{x}}_{1:T}||_{\infty} \leq \epsilon,
\end{equation}
where $||\cdot||_{\infty}$ represents the l-infinity norm. During the gradient-based learning of the adversarial audio segment $\hat{\tilde{\mathbf{x}}}$, this l-infinity norm constraint is incorporated by clamping the values at $\epsilon$~\cite{madry2019deep}. In our experiments (Section \ref{sec:experiments}) we explore the impact of varying these two imperceptibility constraints on the efficacy of the control-attack.

\section{Experiments} \label{sec:experiments}
\subsection{Experimental Setup}

\noindent \textbf{Attack Configuration.} As described in Section \ref{sec:attack}, we learn a universal adversarial acoustic segment that can be prepended to any input speech signal to \textit{control} Whisper to perform speech translation, despite Whisper being deployed in the speech transcription setting. We present experimental results for Whisper medium (769M parameters). Smaller models are not considered as it is found that due to their lower performance in speech translation it limits the potential of the model-control attack (the model-control attack assumes the model has the ability to perform the target task). Following the imperceptibility definitions given in Section \ref{sec:att-method}, we learn universal adversarial acoustic segments of three different strengths: 1) a weak attack (\textit{attack-w}) with the harshest constraints of a maximum amplitude $\epsilon=0.02$ and an audio length of 10,240 frames, equivalent to 0.64 seconds for audio sampled at 16kHz; 2) a mid strength attack (\textit{attack-m}) with $\epsilon=0.2$ and length 0.64 seconds; and 3) a strong attack (\textit{attack-s}) with $\epsilon=2.0$ and length 5.12 seconds.\newline

\noindent \textbf{Data.} To illustrate the impact of the universal model-control attack to force Whisper to \textit{translate} instead of transcribe, we use the popular Few-shot Learning Evaluation of Universal Representation of Speech (FLEURS) dataset~\cite{fleurs2022arxiv}. It is a n-way parallel speech dataset in 102 languages, with 12 hours of speech per language.  We select the French-English language pair as the primary development set, where we present the results and analysis for the \textit{weak}, \textit{mid} and \textit{strong} attacks in Section \ref{sec:main}. To assess the generalizability of the universal attack method to other language pairs, we carry out an ablation study for the following other aligned language pairs: German-English (de-en), Russian-English (ru-en) and Korean-English (ko-en). The train splits of each dataset are used to learn the universal attack, and experimental results are reported on the unseen test splits. As the Whisper model is only trained to perform speech translation from a language X to English, in our experiments language X (where X is not \textit{en)} audio is input to Whisper and the aim of the attack is to encourage Whisper to translate it to English. \newline 

\noindent \textbf{Metrics.} We use standard popular ASR and speech translation metrics to measure the impact of the attack. This includes Word Error Rate (WER); the BLEU score~\cite{Papineni2002BleuAM}; and the COMET score~\cite{rei2020cometneuralframeworkmt}. Finally, as the aim of the attack is to cause Whisper to translate the input audio to English, we also report the average (across the dataset) probability of English, P(en), as given by Google's LangDetect model~\cite{nakatani2010langdetect}.

\subsection{Attack Results and Analysis} \label{sec:main}

The model-control universal acoustic segments are trained such that they can be prepended to any input speech signal and cause Whisper to perform speech translation despite being set to perform speech transcription. Using French-English (fr-en) as the primary language pair for evaluating the model-control attack, Table \ref{tab:fr} presents the impact of the model-control attacks of each strength. For reference, the performances with respect to the English transcriptions are given when Whisper is run with no attack in transcription mode (expected to be low performance as Whisper transcribes in the source audio language French) and when Whisper is run with no attack in translation mode (the upperbound of performance for the attack). As the attack strength is increased from \textit{weak} to \textit{strong}, the attack performance approach the translation mode upperbound performance. This demonstrates that the model-control attack can be successful in overriding the transcription mode to cause Whisper to perform speech translation instead.

\begin{table}[htb!]
    \centering
    \small
    \begin{tabular}{l|c|cccc}
    \toprule
        & Mode & WER$\downarrow$ & BLEU$\uparrow$ & COMET$\uparrow$ & P(en)$\uparrow$\\ \midrule
        No Attack & tc & 115 & 0.42 & 14.9 & 0.0\\
         No Attack & tl & 55.3 & 17.4 & 51.0 & 100\\ 
         Attack-w & tc & 91.5 & 8.63 & 16.5 & 42.9\\
         Attack-m & tc & 79.4 & 11.0 & 32.9 & 62.4\\
         Attack-s & tc & 57.6 & 17.5 & 47.4 & 98.2\\
         \bottomrule
    \end{tabular}
    \caption{Model-control attack on \textbf{fr-en}: universal prepend attack to override Whisper's transcribe (tc) mode and perform speech translation (tl). Input audio is French and metrics evaluated with respect to reference English translations.}
    \label{tab:fr}
\end{table}

\begin{figure}[t]
     \centering
     \begin{subfigure}[b]{0.9\linewidth}
            \centering
        \includegraphics[width=\linewidth]{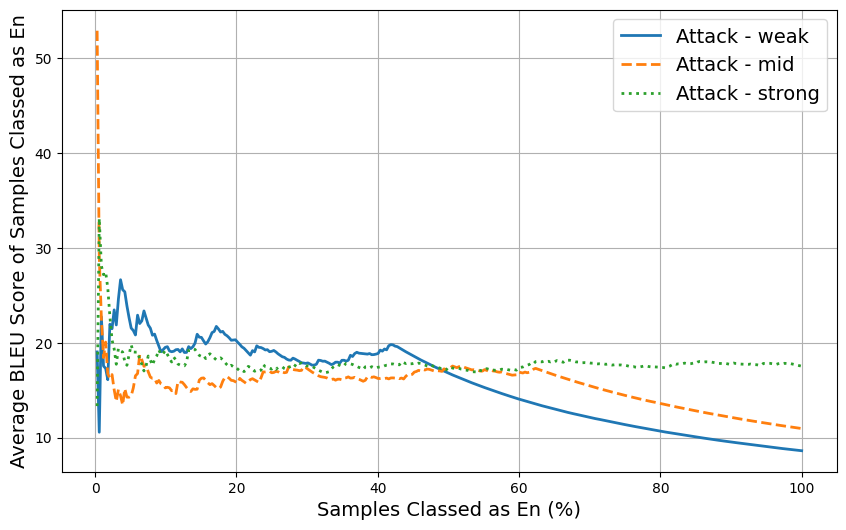}
        \caption{BLEU-recall success}
     \end{subfigure}
     \hfill
     \begin{subfigure}[b]{0.9\linewidth}
            \centering
        \includegraphics[width=\linewidth]{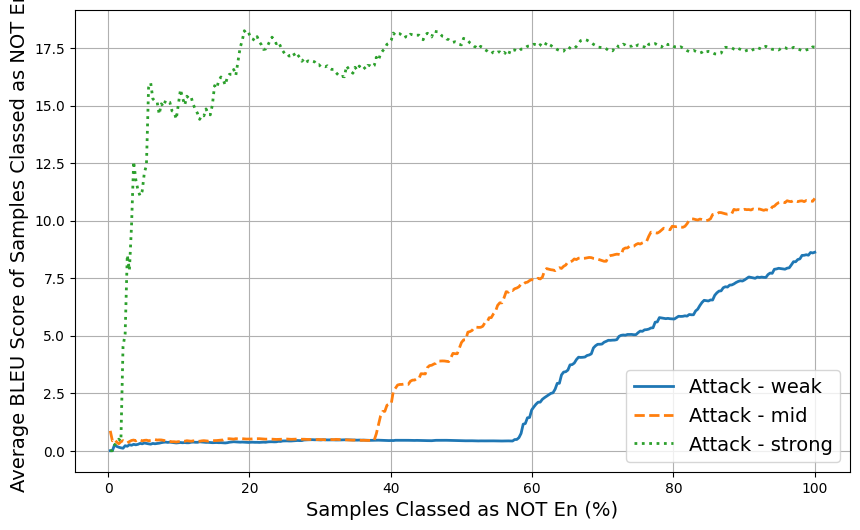}
        \caption{BLEU-recall fail}
     \end{subfigure}
     \hfill
    \caption{BLEU performance for \textit{recalled} samples where samples are recalled if the model-control attack is successful or fails, as per the P(en). Curves for fr-en data samples.}
        \label{fig:recall}
\end{figure}

Although the model-control attack is highly effective in forcing Whisper to perform speech translation, it is unable to perfectly match the upperbound performance with Whisper running freely in translation mode (Table \ref{tab:fr}). This could be due to either the attack failing to make Whisper enter \textit{translation} mode for specific input speech signals or due to the attack resulting in Whisper generating lower quality translations than when Whisper is run directly in translation mode. We next investigate the cause of the discrepancy. For each attack strength, we compute \textit{BLEU-recall} curves for samples on which the universal model-control attack is `successful' and `unsuccessful' (Figure \ref{fig:recall})~\footnote{Identical trends were observed with \textit{COMET}-recall curves.}. A sample is classified as successfully attacked if \(P(\text{en}) > \tau\). Conversely, a sample is classified as a `failed' attack if \(P(\text{en}) < \tau\). We sweep the arbitrary threshold \(\tau\) from 0 to 100\% and compute BLEU scores for both the successfully and unsuccessfully attacked samples \textit{recalled} at each value of $\tau$. These curves help explain the quality of the translations generated by the model when attacked, separately for when the attack succeeds or fails to make Whisper generate English tokens (measured by P(en)). 

\begin{figure}[htb!]
    \centering
    \includegraphics[width=0.9\linewidth]{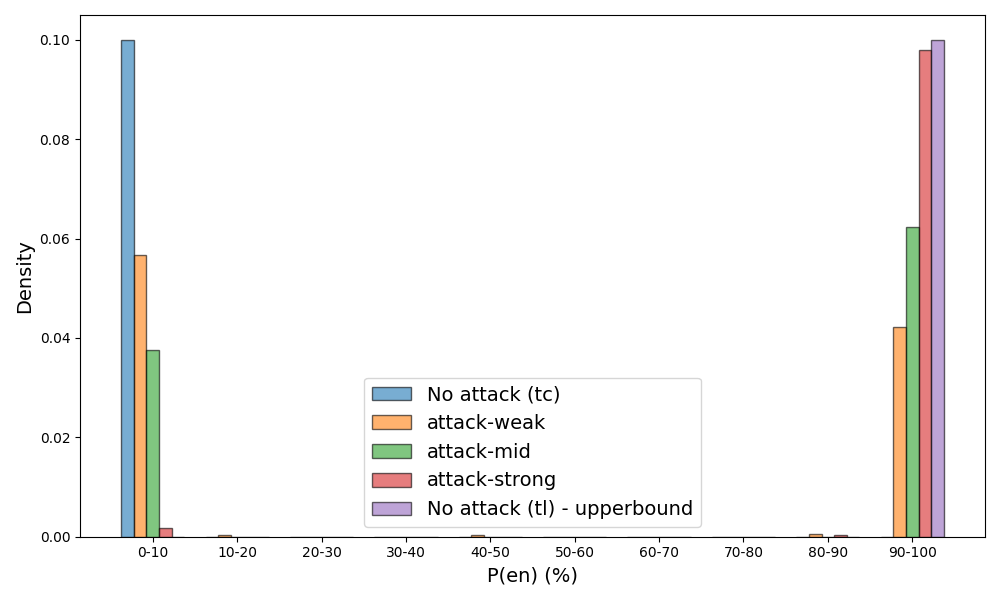}
    \caption{P(en) (\%) distribution}
    \label{fig:pen}
\end{figure}

From \ref{fig:recall}, in general the greater the probability of English in the generated translation, the higher the quality of the translation as measured by BLEU. However, the curves also reveal a more interesting behaviour of the attacks: the success recall curves (\ref{fig:recall}a), have an observable discontinuity in the BLEU scores when 43\%, 62\% and 98\% of the samples are `successfully' attacked for the weak, mid and strong attacks respectively, and conversely in the fail recall curves (\ref{fig:recall}b), there is a discontinuity when approximately 57\%, 38\% and 2\% samples `fail' to be attacked. These percentages almost perfectly align with the average probability of English given by each attack in Table \ref{tab:fr}. This suggests that the attacks result in a highly bi-modal response, where the attack on a specific input audio sample is either perfectly successful in causing Whisper to generate English text or perfectly unsuccessful, with near 0\% probability of English. This bi-modal split is further verified in the distribution of P(en) for each attack in Figure \ref{fig:pen}. With Whisper in transcribe (tc) mode, it is entirely non-English (left-most bar), whilst when Whisper in translate (tl) mode, it generates entirely English text (right-most bar). However for all of the attacks in Figure \ref{fig:pen}, we observe the above-mentioned bi-modal pattern, where the model-control attacks do not result in a gradual distributional shift from the French transcriptions to English translations, but instead display binary success. The stronger the attack the greater the proportion of the samples that are `successfully' attacked (greater fraction of samples on the right side) resulting in almost complete English text. On the whole, the universal model-control attack, when successful, causes Whisper to behave as though as if it is in translation mode, but when the attack is unsuccessful it results in Whisper continuing to behave as set in its default transcription mode - there are no partial modes. It is the strictness around the imperceptibility constraints (strength of the attack) that dictates the fraction of samples the universal model-control attack is successful upon.

\subsection{Language Ablation} \label{sec:ablation}

In this section we assess the portability of the model-control attack method to other language pairs beyond French-English, where the aim as before is to learn a universal acoustic segment that can be universally prepended to any input speech signal and cause Whisper to carry out speech translation (to English), despite being deployed in speech transcription mode. We consider the \textit{strong} attack setting in this section.
\begin{table}[htb!]
    \centering
    \small
    \begin{tabular}{l|c|cccc}
    \toprule
        & Mode & WER$\downarrow$ & BLEU$\uparrow$ & COMET$\uparrow$ & P(en)$\uparrow$\\ \midrule
        \multicolumn{6}{c}{{\bf French-English (fr-en)}}\\
        \midrule
        No Attack & tc & 115 & 0.42 & 14.9 & 0.0\\
         No Attack & tl & 55.3 & 17.4 & 51.0 & 100\\ 
         Attack & tc & 57.6 & 17.5 & 47.4 & 98.2\\
         \midrule
         \multicolumn{6}{c}{{\bf German-English (de-en)}}\\
         \midrule
        No Attack & tc & 101 & 0.29 & 3.24 & 0.0\\
         No Attack & tl & 53.7 & 18.9 & 45.0 & 99.9 \\ 
         Attack & tc & 78.6 & 13.0 & 23.8 & 95.5\\
         \midrule        
         \multicolumn{6}{c}{{\bf Russian-English (ru-en)}}\\
         \midrule
         No Attack & tc & 101 & 0.10 & 0.16 & 0.0\\
         No Attack & tl & 61.6 & 14.6 & 40.6 & 99.9 \\ 
         Attack & tc & 96.8 & 10.4 & 16.8 & 95.4\\
\midrule
\multicolumn{6}{c}{{\bf Korean-English (ko-en)}}\\
\midrule
        No Attack & tc & 99.8 & 0.00 & 0.26 & 0.3\\
        No Attack & tl & 72.8 & 9.68 & 34.9 & 98.7 \\ 
         Attack & tc & 93.6 & 8.46 & 18.4 & 98.1\\
         \bottomrule
    \end{tabular}
    \caption{Model-control attack (\textit{strong}) on \textbf{fr/de/ru/ko-en}: universal prepend attack to override Whisper's transcribe \textit{tc} mode and perform speech translation (tl). Input audio is in the source language and metrics evaluated wrt to reference English translations.}
    \label{tab:all}
\end{table}

Table \ref{tab:all} presents the impact of the universal model-control attack for the new languages: German-English (de-en), Russian-English (ru-en) and Korean-English (ko-ru). On the whole, the attack appears to fairly successful for all language pairs in forcing Whisper to generate English translations, as indicated by the average probability of English (P(en)) going from 0.0\% to above 95\% for all language pairs. However, when considering the BLEU and COMET scores, it appears the quality of translations with the attack (relative to the upperbound performance given by \textit{no attack} in \textit{tl} mode) for de/ru/ko is not quite as high as for fr, where the \textit{strong} attack closely approached the upperbound performance (Table \ref{tab:fr}). The gap between the attack (\textit{strong}) performance and the upperbound performance for de/ru/ko can be explained by the decomposition of the WER between the translations generated. Table \ref{tab:wer-breakdown} presents the WER decomposition between translations generated by the attack (on Whisper in transcribe model) and the \textit{no attack} translations. It is interesting to note that a significant contributor to the increase in WER for de/ru/ko relative to fr, comes from a large increase in the insertion rate. This suggests that the attack causes Whisper to hallucinate. As an example, we find that for Russian, the attack causes 167 samples to have the phrase \textit{however, it is clear that} inserted at the beginning of the translation, whereas only 1 sample has this phrase in its translation when there is no attack, i.e., it seems that the attack can sometimes can learn an acoustic realization for the prepended acoustic audio, i.e, for some languages the translations generated at inference (free-running) can contain a prepended text sequence before the actual translation, resulting in lower translation quality.
\begin{table}[t]
    \centering
    \small
    \begin{tabular}{l|ccc|c}
    \toprule
        Lang & ins & del & sub & WER \\ \midrule
        fr-en & 6.7 & 4.2 & 10.4 & 21.3\\
        de-en & 18.9 & 14.3 & 18.3 & 51.3\\
        ru-en & 32.7 & 10.7 & 12.5 & 55.9\\
        ko-en & 31.7 & 10.9 & 21.5 & 64.1\\
        \bottomrule
    \end{tabular}
    \caption{Word Error Rate (insertions, deletions and substitutions) between \textit{attack-strong} on Whisper running in transcribe mode and \textit{no attack} with Whisper running in translate mode.}
    \label{tab:wer-breakdown}
\end{table}

\begin{figure}[t]
     \centering
     \begin{subfigure}[b]{0.9\linewidth}
            \centering
        \includegraphics[width=\linewidth]{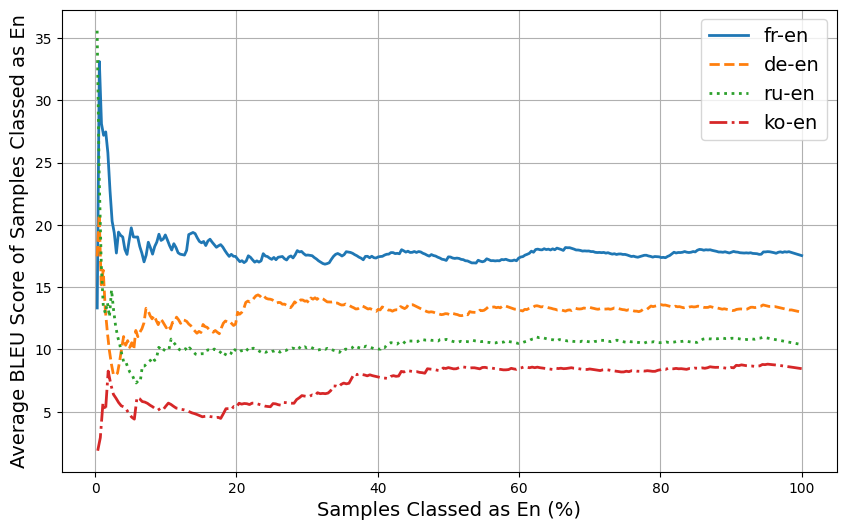}
        \caption{BLEU-recall success}
     \end{subfigure}
     \hfill
     \begin{subfigure}[b]{0.9\linewidth}
            \centering
        \includegraphics[width=\linewidth]{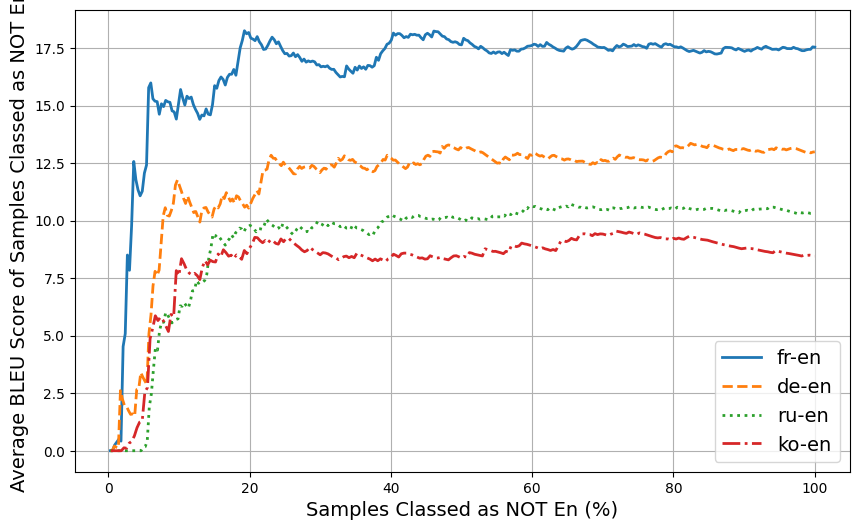}
        \caption{BLEU-recall fail}
     \end{subfigure}
     \hfill
    \caption{BLEU performance for \textit{recalled} samples where samples are recalled if the model-control attack (\textit{strong}) is successful or fails, as per P(en).}
        \label{fig:recall-all}
\end{figure}

Next, to verify that the attacks on de/ru/ko display the same bi-modal behaviour as the attack on fr-en, we consider the BLEU-recall curves for each language pair in Figure \ref{fig:recall-all}. Most easily observable in Figure \ref{fig:recall-all}b the discontinuity in BLEU performance occurs as before with fr-en at the percentage of samples corresponding to approximately the average probability of not English (in Table \ref{tab:all}) - between 2\% and 5\% for the different languages. This verifies that for all languages the same trend holds: the model-control attack is highly binary.

\section{Conclusion}

This work reveals a vulnerability in multi-tasking speech-enabled foundation models, specifically through model-control adversarial attacks. We demonstrate that it is possible to manipulate the task setting of such multi-tasking speech-enabled models. By appending a short universal adversarial acoustic segment to any input speech signal, we were able to override the model's prompt setting and force it to perform speech translation instead of its default speech transcription. An intriguing aspect is the binary nature of the attack's success. The universal adversarial acoustic segment either successfully manipulates Whisper to operate in its translation mode, or it fails entirely, resulting in Whisper behaving in its default transcription mode. This bi-modal pattern indicates that the attack does not create intermediate operational states, but rather a clear switch between transcription and translation. The strictness around the imperceptibility constraints dictates the fraction of samples the universal model-control attack switches. The risk of model-control attacks highlights the need for increased security measures in the deployment of flexible ASR systems. As these models are designed to perform an increasing number of tasks within a single framework, they become more susceptible to such adversarial manipulations. We encourage future work on developing robust defenses against model-control adversarial attacks.~\footnote{Acknowledgement: We thank Cambridge University Press \& Assessment for their support in funding this work.}

\newpage

\bibliographystyle{IEEEbib}
\bibliography{manual}

\end{document}